\documentclass[twocolumn,showpacs]{revtex4}

\usepackage[latin1]{inputenc}
\usepackage[dvips]{graphicx}
\usepackage{color}
\usepackage{amssymb}
\usepackage{amsmath}
\usepackage{hyperref}
\usepackage{subfigure}

\begin{document}

\title{A discrete statistical mechanics approach to aeolian ripple dynamics}

\author{
Franco Bagnoli$^{1,3}$\thanks{franco.bagnoli@unifi.it},
Duccio Fanelli$^{1,2,3}$\thanks{duccio.fanelli@cmb.ki.se},
Luca Sguanci$^{1,3}$\thanks{luca.sguanci@unifi.it}
}

\affiliation{
1. Dept. of Energy, University of Florence,
via S. Marta, 3, 50139 Firenze, Italy\\
2. Department of Cell and Molecular Biology, Karolinska Institute,
SE-171 77 Stockholm, Sweden\\
3. INFN, sect. Florence and Center for the Study of Complex Dynamics, University of Florence. 
}

\date{\today}

\begin{abstract}
The interaction between a fluid and a granular material plays a   crucial role in a large class of  phenomena such as landscape   morphology and transport of sediments, aeolian sand dunes   formation and ripples dynamics. Standard models involve deterministic continuum  equations or, alternatively, Lattice Boltzmann and Lattice Gas Cellular Automata. We here introduce a toy-model to address the issue of aeolian ripple formation and evolution. Our simplified approach accounts for the basic physical mechanisms and enables to   reproduce the observed phenomenology in the framework of an  innovative statistical mechanics formulation.
\end{abstract}


\maketitle

\section{Introduction and general background}\label{intro}

The complex interactions between a granular material and  a fluid may lead to the formation of spectacular structures, like erosion patterns, dunes and ripples. We are here interested in ripple patterns that emerge due to the interactions between a fluid (air) flow and a granular material (sand). The conditions of the flow and the characteristics of the granular material give origin to a large variety of morphologies~\cite{Bagnold,Pye}. 

\begin{figure}
\begin{center}
\includegraphics[width=0.6\columnwidth]{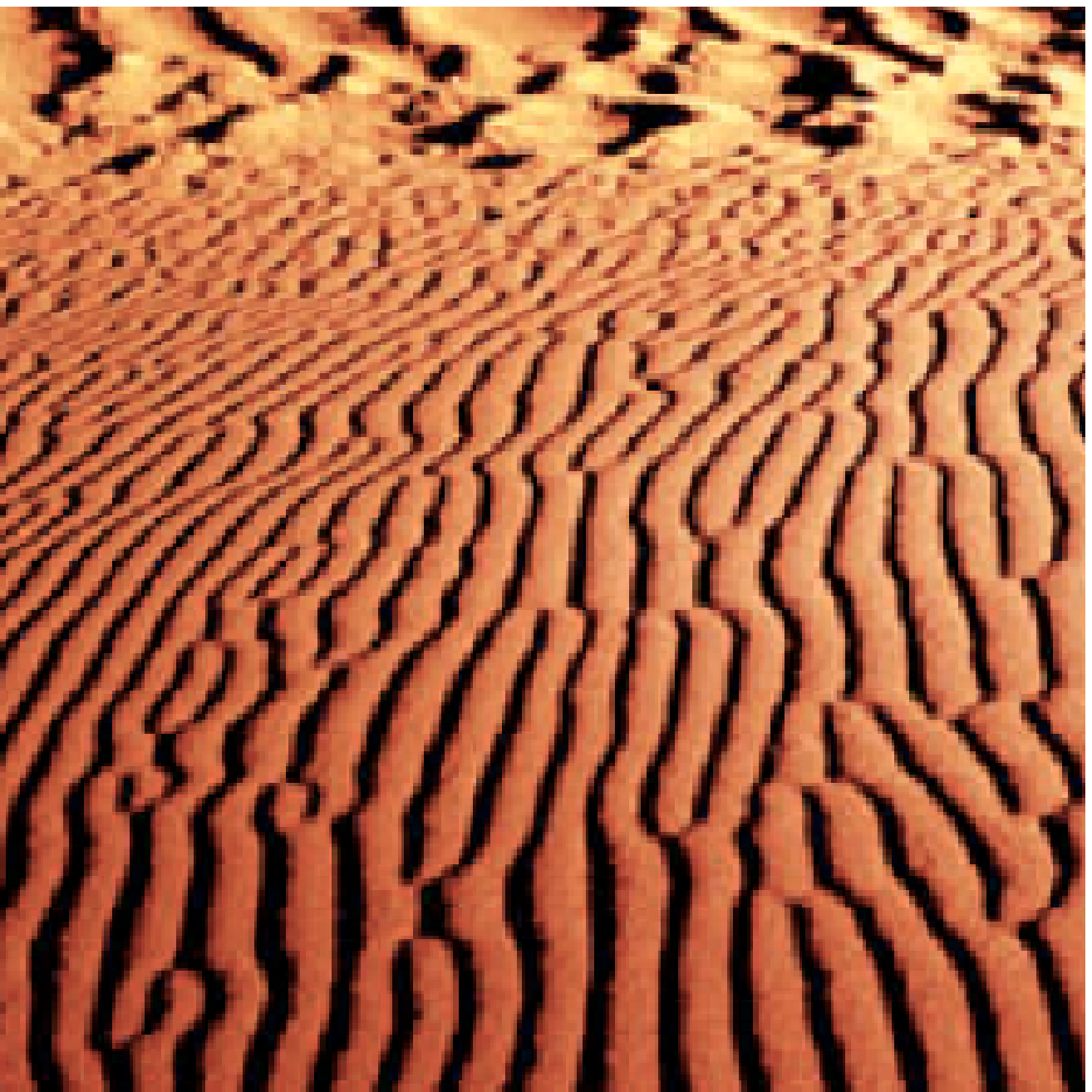} 
\includegraphics[width=0.6\columnwidth]{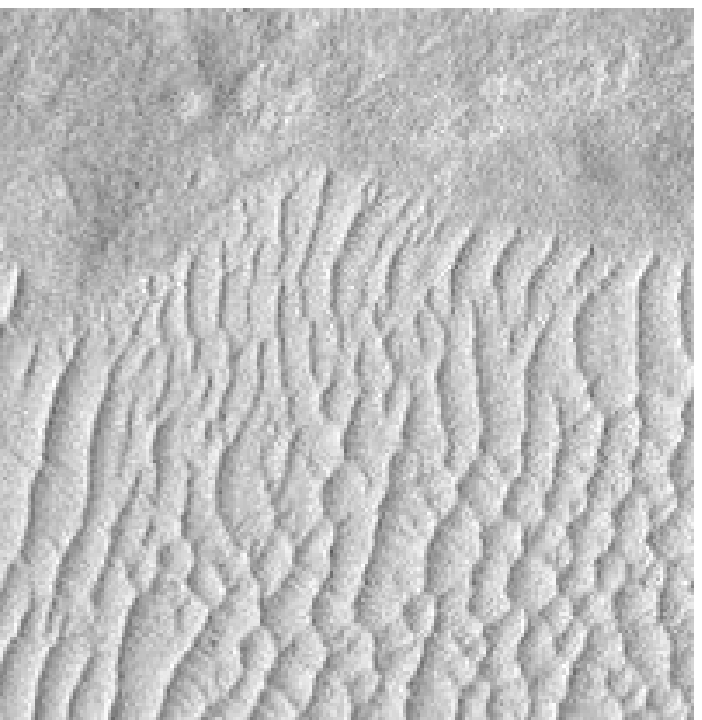} 
\caption{(color online) Pattern of aeolian ripples in south central Colorado (upper)~\protect\cite{nature}. 
Giant sand dunes and ripples, 
steeper than on Earth, has been recently observed on Mars (NASA)
(lower)~\protect\cite{nasa}.}
\label{fig:ripples}
\end{center}
\end{figure}

Ripples have been recently photographed even on Mars, qualitatively resembling the ones observed on the Earth surface, see Fig.~\ref{fig:ripples}.  From a direct inspection of the peculiar shapes of such coherent structures, one could in principle aim at reconstructing  the characteristics of the surrounding environment.  It is therefore of paramount importance to develop dedicated models enabling to successfully address the  issue of ripple formation and evolution. Phenomenologically, the correct interpretative framework is provided by the pioneering investigations by Bagnold~\cite{Bagnold} based on wide observational and experimental facts. In particular,  the two key processes, namely  \emph{saltation} and surface creeping (\emph{reptation}), are outlined in Ref.~\cite{Bagnold}.
 
Saltation refers to the process by which a  grain is entrained by the fluid, accelerated  and transported to another site. In wind flows the transition from rest to entrainment  occurs as a sudden jump, when the lift force is able to dislodge a particle from its gravitational trap. Due to the large density differences between sand and air, this occurs for large shear velocities. The takeoff angle (with respect to horizontal) is large,  and so is the flight time and transport. The subsequent impact with the other particles occurs at relatively high energy, and even when the restitution coefficient is small, this energy may be sufficient to dislodge other particles. These \emph{reptated}  particles are then transported by the fluid, but their initial velocity is generally  lower than that associated to the saltating particles, thus preventing  avalanche effects.

Reptated particles finally stop in a local gravitational trap, loosing energy via anelastic collisions. One of the standard assumption~\cite{NO1993} is that this reptation mechanism is the main origin of the creeping flow, including what in  other context would be referred as toppling.

Which one of these phenomena is more relevant, depends on the combined effects of drag and lift forces, gravity and dissipation of energy during collisions. However the main difficulty arises in modeling the interaction between the fluid and the grains. This interaction depends on the distribution of the  fluid velocity, that may vary widely in space and time, on the shape of the grains and on their relative density, that gives origin to buoyancy.  Moreover, the instantaneous fluid velocity field depends on the profile of the sand bed, which is modified by erosion and deposition, and on the shielding effect of saltating particles. Self-organized patterns are originated by these nontrivial interactions~\cite{Andreotti,KH2002}.  
Dedicated models aim at reproducing some of the most relevant aspects of these phenomena. In particular, following Ref.~\cite{Bagnold},   ripples are characterized by being asymmetric and their formation and persistence is determined by flux intensity. Small ripples are observed to travel faster than large ones and the temporal evolution of the maximum ripple  height is limited and not linear.

In order to study the erosion/deposition process continuum models of mechanics have been proposed~\cite{TB1998,VR1999,LP1999}. 
In addition accurate hydrodynamical descriptions have been studied by various authors~\cite{KH2002,AV2005,PL2003,CV2000}.
These alternative approaches enable to quantitavely study the process of ripple formation by incorporating a detailed representation of the fluid flow via Navier-Stokes equation, and/or facing the problem of considering the flow near the soil.  However, the latter models are computationally expensive and difficult to treat analitically particularly in presence of complex and time-dependent boundaries, or when turbulence and other fluctuating aspects play an active role. However, linear stability analysis~\cite{TB1998,VR1999,Andreotti} enables to predict the instability regime of a flat surface and quantifies its associated growth rate. 

To gain more insight into the crucial interplay between erosion and deposition, beyond the linear approximation, a series of simplified theoretical frameworks (toy models) have been developed, focusing only on those aspects supposed to be the relevant ones.  
Following these lines, it is customary to replace the complex fluid velocity distribution with a limited number of aggregated data, like the average shear velocity and the associated fluctuations. 

Simple models of sand ripples dynamics were first introduced by Anderson~\cite{RA1987,RA1990} for grain segregation and stratigraphy. A discrete stochastic model was further proposed about a decade ago by Werner and Gillespie~\cite{WG1993}. Another minimal model, widely adopted in the relevant literature, was proposed the same year by Nishimori and Ouchi~\cite{NO1993}.
The latter, termed NO, qualitatively captures some of the peculiar features of ripple evolution. Within this scenario, the saltation and reptation are accounted for and shown to produce the spontaneous formation of a characteristic ripple patterns.   Though it represents a significant step forward in the comprehension of the basic mechanism underlying the phenomenon, the NO model allows for non realistic structures of infinite heights, since in this model there is no explicit or implicit mechanism that leads to the appearance  of a critical angle of repose.  Recently,  Caps and Vandewalle~\cite{Caps} modified the preexisting scheme by including explicitly the effect of avalanches (SCA model: Saltation Creep and Avalanches). This modification results in asymmetric ripple profiles and induces a saturation for the maximal height.

Cellular automata modeling of sand transportation was introduced by 
Anderson and Bunnan~\cite{AB1993} and by Werner and Gillespie~\cite{WG1993}.
In these models, the driving mechanism for sand transport is the saltation/reptation dynamics, eventually complemented by  toppling, that corresponds to diffusion in a continuous model.  Masselot and Chopard~\cite{chopard} also introduced a cellular automata for snow and sand transportation.  They explicitly modeled the fluid flow by means of lattice Boltzmann methods, while the granular phase is represented as a probabilistic cellular automaton. The erosion mechanism here is modeled by a constant probability of detachment, and local rearrangements are again  achieved by a toppling mechanism. 

The elementary building blocks of these stochastic models, like the erosion, deposition and toppling steps, have a phenomenological nature, implying that the probability of their occurrence has to be measured experimentally. In this work we propose to adopt a standard statistical mechanics description of the elementary steps, coupled with an external forcing that drives the system out of equilibrium. The role of temperature is here played by the fluctuations of the velocity field.

We focus on the formation and evolution of aeolian ripples. 
The proposed local-equilibrium dynamics enables to reproduce the main characteristics of ripple dynamics, like the observed stable states and the saturation of ripples heights, without including an explicit mechanism for toppling or other local rearrangements. This simple scheme may be easily extended to include a more detailed description of a real fluid. 

The paper is organized as follows: in section~\ref{sec:model} the model is
introduced. Section~\ref{sec:numimp} is devoted to discuss the numerical
implementation. Results are presented in Section~\ref{sec:results}. Finally
in Section~\ref{sec:conclusion} we draw our conclusion.

\section{The model}
\label{sec:model}

\begin{figure}
\begin{center}
\includegraphics[width=1.0\columnwidth]{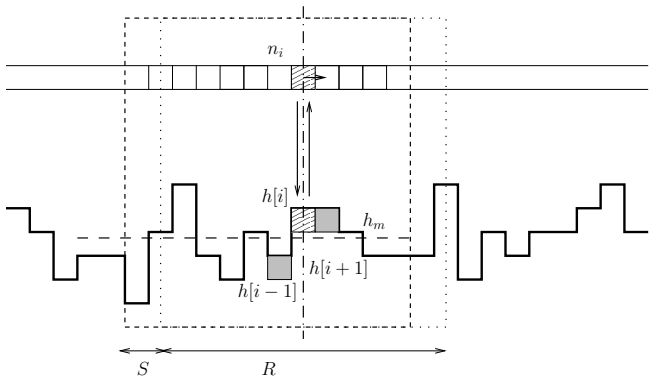} 
\caption{Sand bed schematization in the 1D model. For a given site $i$
of height $h_i$, the symmetric interval of neighbors of amplitude $R$
is shown (dotted line). In order to account for flux asymmetry the
interval considered is shifted of an amount $S$ (dashed line).}
\label{fig:profilo}
\end{center}
\end{figure}

We consider a one dimensional discrete model, $L$ being the extension
of the segment partitioned in $N$ equally spaced intervals, 
and assuming periodic boundary conditions, Fig.~\ref{fig:profilo}.

To each site $i$ we associate the height $h_i$ which labels the number
of particles constituting the $i^{th}$ slice of the \emph{sand bed}.
Further, we consider a bunch of $n_i$ particles flowing over
the bed. The system is therefore composed of two interacting layers
of particles. 

The two processes driving the dynamics of the system are:
\emph{erosion}, which occurs when a resting grain belonging to the
surface of the sediments layer is  entrained by the fluid and
\emph{deposition}, that mimics the deposition of a flowing grain.

The evolution of the system is then modeled as follows: focusing on
the $i^{th}$ site, we select an interval of  $R$ neighbors,
asymmetrically shifted by an amount $S$, see Fig.~\ref{fig:profilo}.
The  amplitude of the interval accounts for the range of local
interactions and is shown to be correlated to the shape of the
ripple.  The quantity $S$ is introduced to model hydrodynamics
effects, such as lift and drag forces~\cite{Graf,Andreotti,KH2002},
which   result in an asymmetry of the flow.

\begin{figure}
  \begin{center}
    \includegraphics[width=1.0\columnwidth]{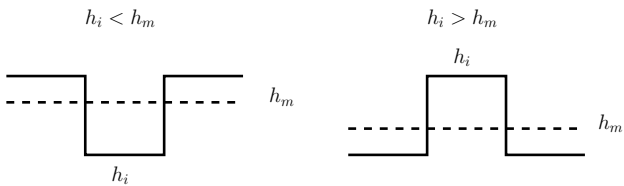}
    \caption{In order to account for bed geometry and hydrodynamics
    effects, we suppose a grain on the bed surface to experience a force
    proportional to the difference between its height and the mean height
    of a local interval of neighbors. In particular, the erosion
    (deposition) probability increases as the height of the site being
    considered is higher (lower) than the mean height of the neighboring
    sites.\label{fig:hm}}
  \end{center}
\end{figure}

We then calculate the mean height of the selected interval, $h_m$. As
a reasonable hypothesis, assume that the probability to experience an
erosion event increases by augmenting the gap between $h_i$ and $h_m$,
under the constraint $h_i>h_m$. Conversely, the deposition will 
most probably occur when the positive difference $h_m-h_i$ gets larger, 
Fig.~\ref{fig:hm}. 

Following the previous reasoning, as a first approximation, we suppose
the force acting on a particle to depend on the difference between the
height of the site being considered and the mean height of the
neighbors.  The energy scales therefore as:
\begin{equation}
  E_i = (h_m-h_i)^2~.
  \label{eq:energy}
\end{equation}
The erosion and deposition processes are hence characterized by means
of the following change in energy:
\begin{center}  
  \begin{tabular}{|c|c|}
    \hline
    \textbf{Erosion}    & \textbf{Deposition}\\
    \hline
    $h_i(t+1)=h_i(t)-1$  & $h_i(t+1)=h_i(t)+1$ \\
    $\Delta
      E_{er,i}=2(h_m-h_i+0.5)$ &  $\Delta
      E_{dep,i}=2(h_i-h_m+0.5)$ \\ 
      \hline
  \end{tabular}
\end{center}

Consequently, it is reasonable to assume the erosion and deposition
probabilities~\cite{Metropolis}: 
\[
 \begin{split}
  P_{\text{er},i} &= \begin{cases}1&\text{if $\Delta E_{\text{er},i}<0$,}\\
   \exp(-\beta_{e} \Delta E_{\text{er},i}) &\text{otherwise;}\end{cases}\\
  	P_{\text{dep},i} &= \begin{cases}1&\text{if $\Delta E_{\text{dep},i}<0$,}\\
  \exp(-\beta_{d} \Delta E_{\text{dep},i})&\text{otherwise.}\end{cases}
 \end{split}
\]
where $\beta_{e}$ and $\beta_{d}$ are constant parameters, analogous
to the inverse of effective temperatures $1/T_{e}$, $1/T_{d}$. The
system is largely out of equilibrium, and these temperatures are
assumed to be related to the amplitude of the fluctuations of the
velocity field in correspondence of the typical erosion and deposition
events.

\section{Numerical implementation} \label{sec:numimp}
The evolution rule, applied simultaneously to each site, is based on a
Metropolis Monte Carlo dynamics~\cite{Metropolis}: first the
deposition is made to happen followed by the subsequent erosion step.
Focusing on the $i^{th}$ site, the deposition step yields:
\begin{enumerate}
  \item for each of the $n_i$ flowing particles a uniformly
    distributed random number $r$ is extracted;
  \item if $r < P_{d}$ deposition occurs and the height of the site is
    increased by one, $h_i(t+1)=h_i(t)+1$, while the bunch of flowing
    particles is decreased by one, $n_i(t+1)=n_i(t)-1$,
\end{enumerate}
The erosion is characterized by:
\begin{enumerate}
  \item a uniformly distributed in the unit interval random number $r$
    is generated;
  \item according to the Monte Carlo Metropolis rule the site is
    eroded if $r < P_{e}$; in this case the height of the site is
    decreased by one, $h_i(t+1)=h_i(t)-1$, and the pool of eroded particle
    is increased by one, $n_i(t+1)=n_i(t)+1$.
\end{enumerate}

The procedure is iterated and the evolution of the heights monitored.

\begin{figure}
  \begin{center}
    \includegraphics[width=0.7\columnwidth]{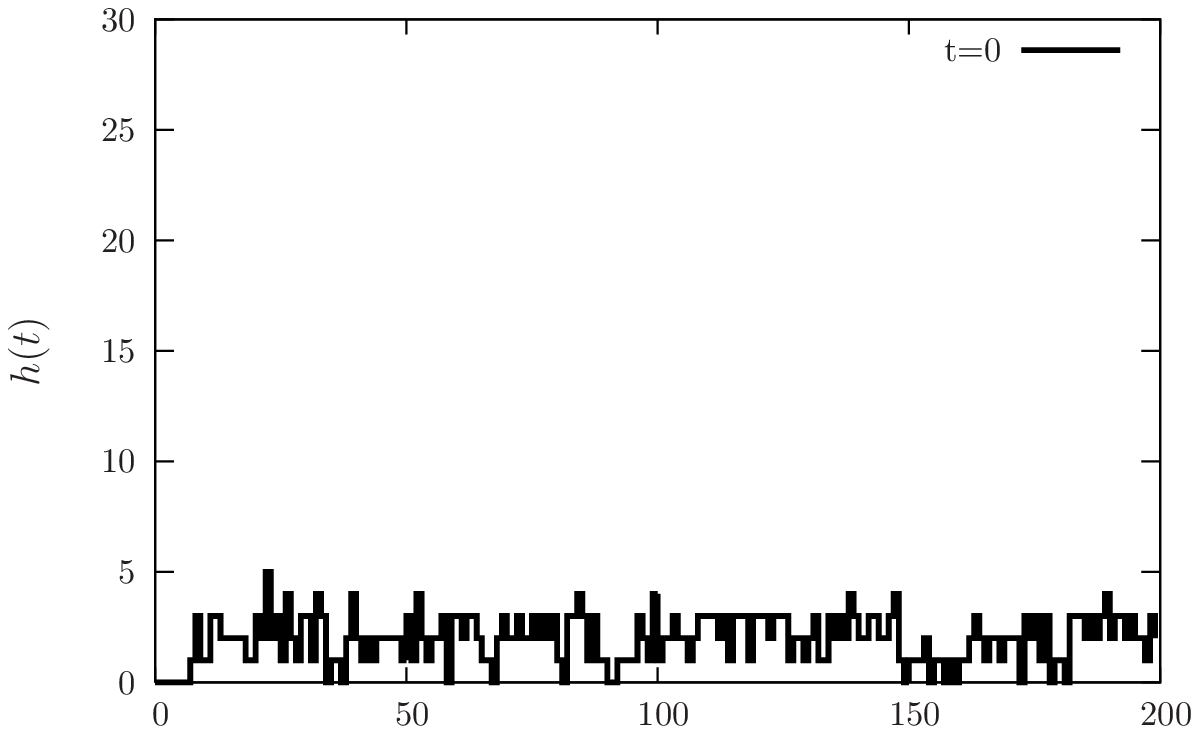}
    \includegraphics[width=0.7\columnwidth]{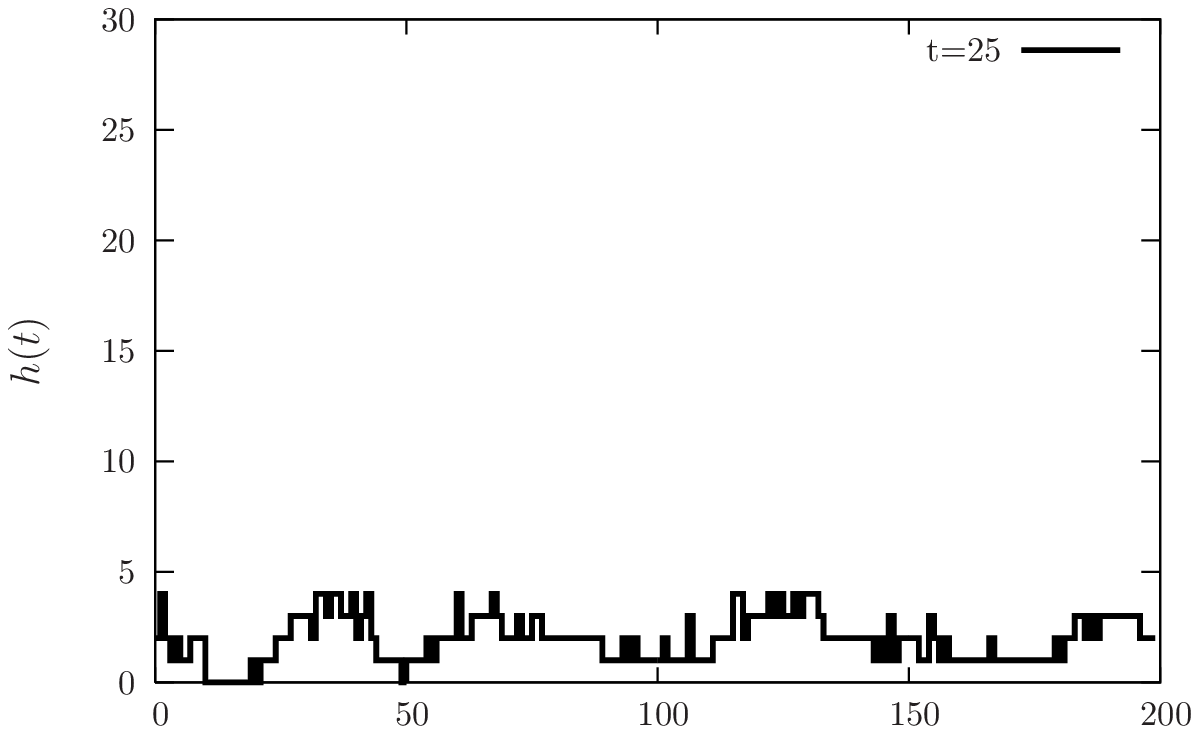}\\
    \includegraphics[width=0.7\columnwidth]{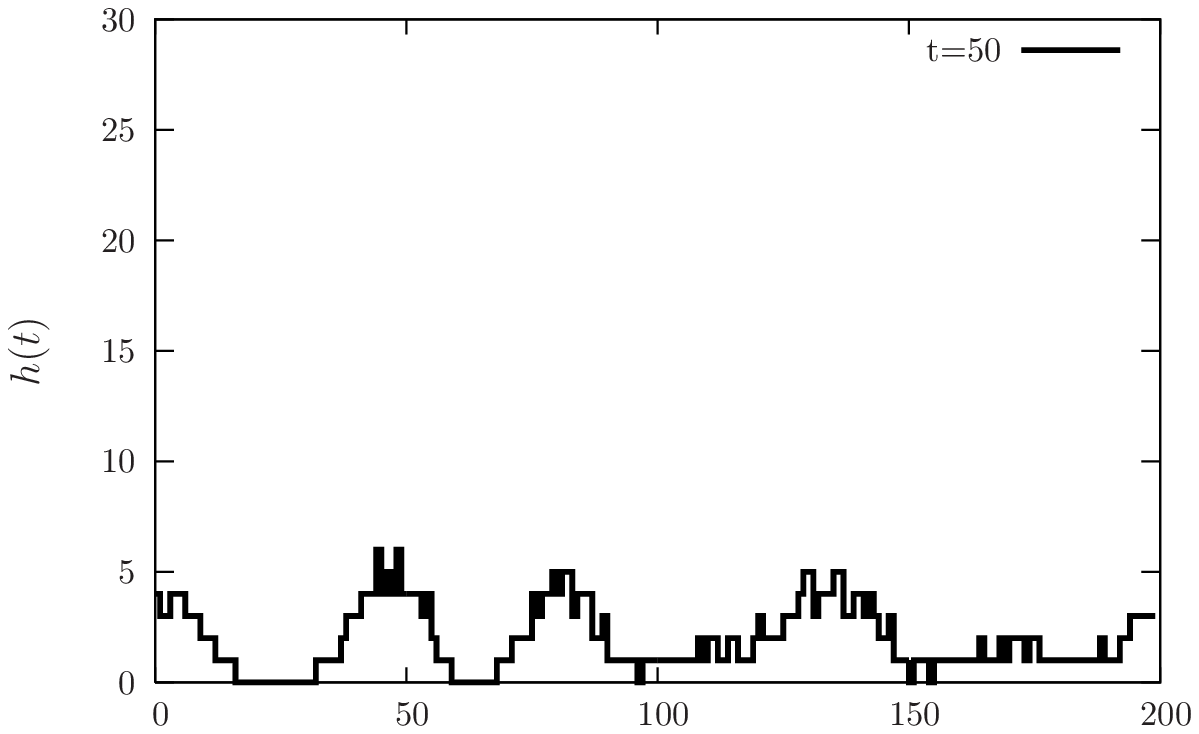}
    \includegraphics[width=0.7\columnwidth]{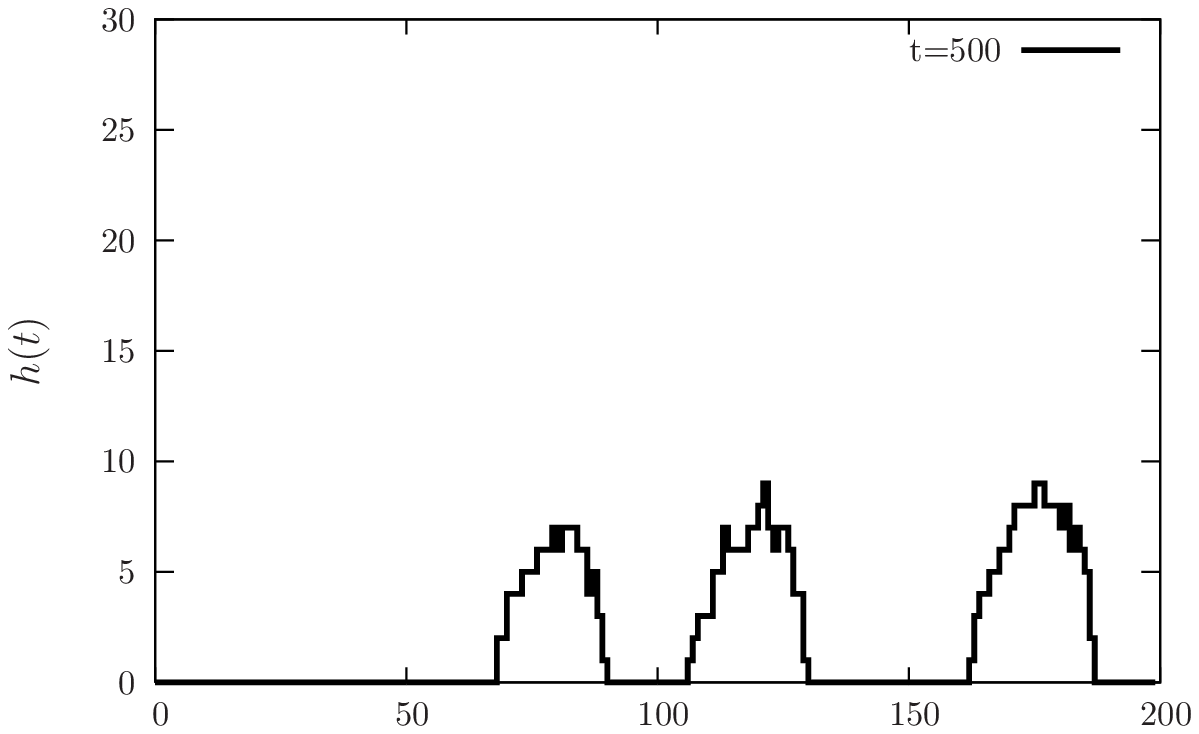}\\
    \includegraphics[width=0.7\columnwidth]{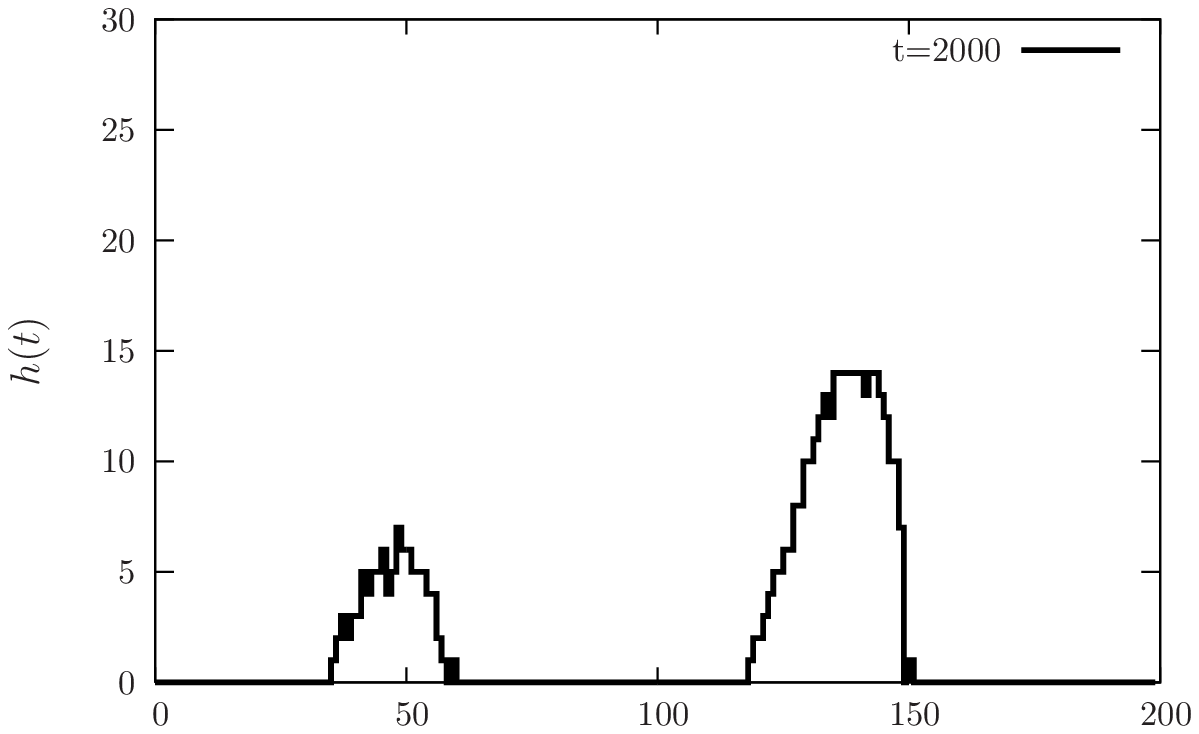}
    \includegraphics[width=0.7\columnwidth]{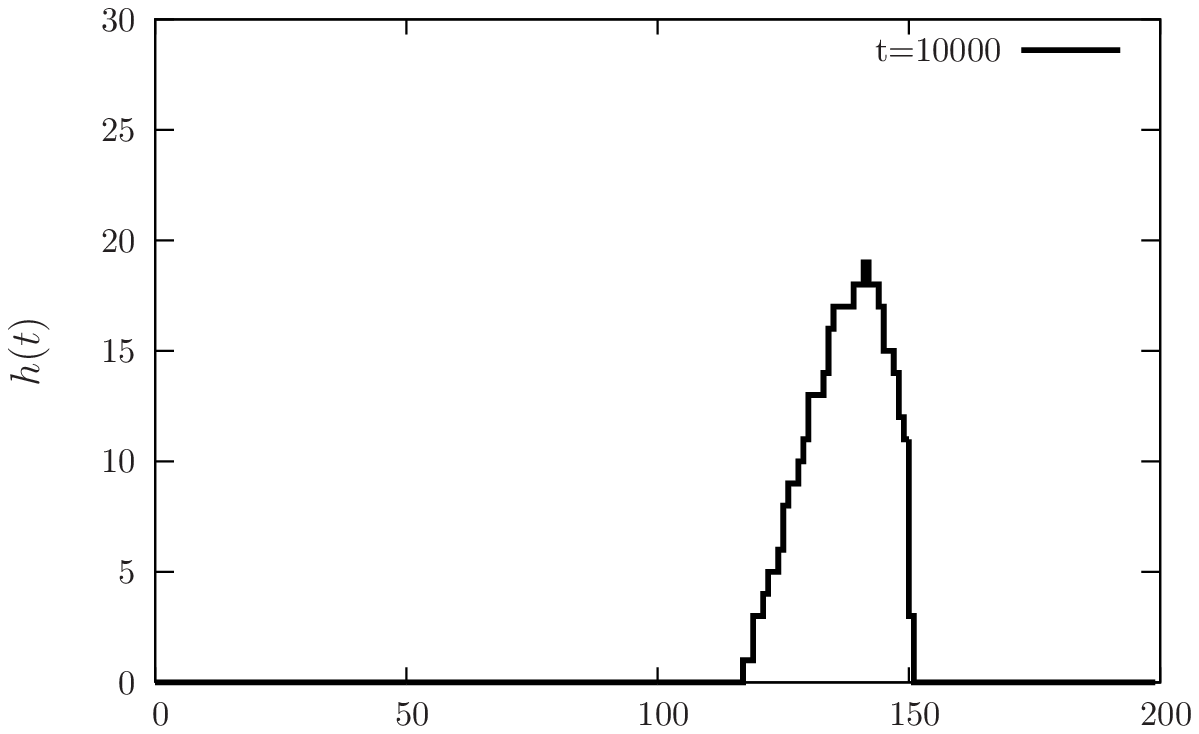}
    \caption{Time snapshots of ripple dynamics. Note that the scales
    of the two axes are different.\label{fig:evolution}}
  \end{center}
\end{figure}

\section{Results}\label{sec:results}
Numerical simulations are performed starting from an initial uniformly
random generated river-bed. Small inhomogeneities are enhanced 
as time progresses, and eventually result in macroscopic ripples 
that display a characteristic asymmetric profile.

\begin{figure}
  \begin{center}
    \includegraphics[width=0.7\columnwidth]{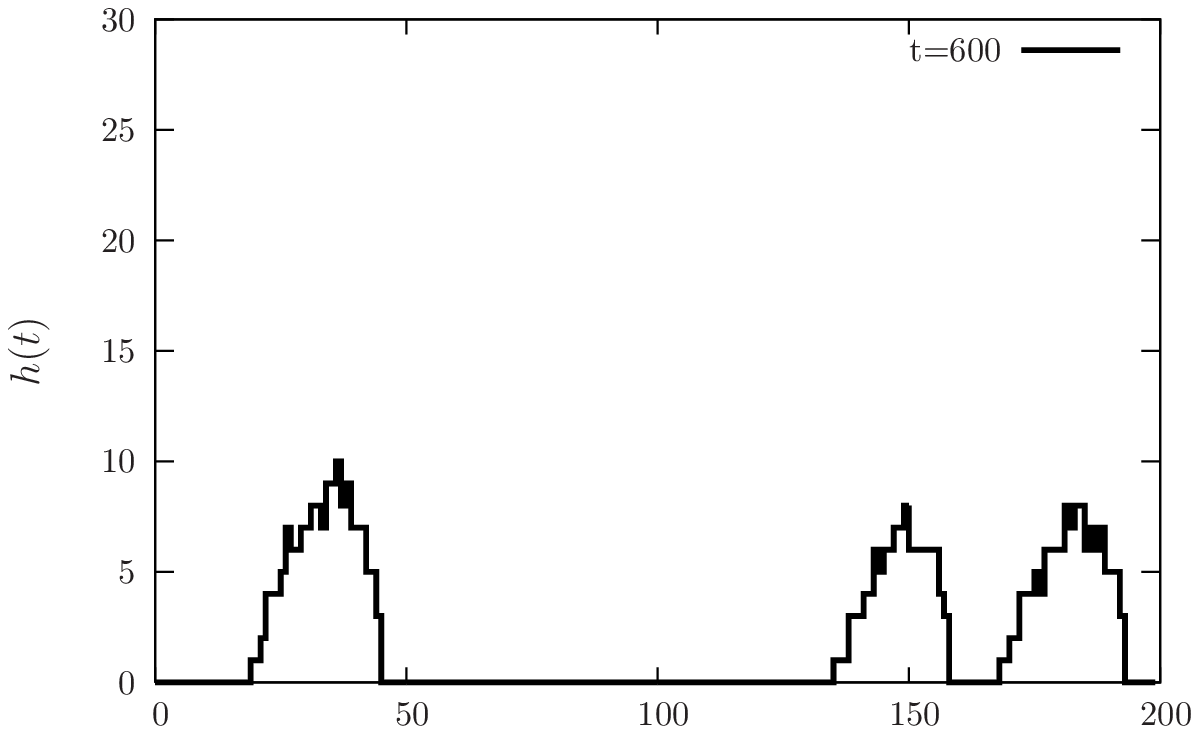}
    \includegraphics[width=0.7\columnwidth]{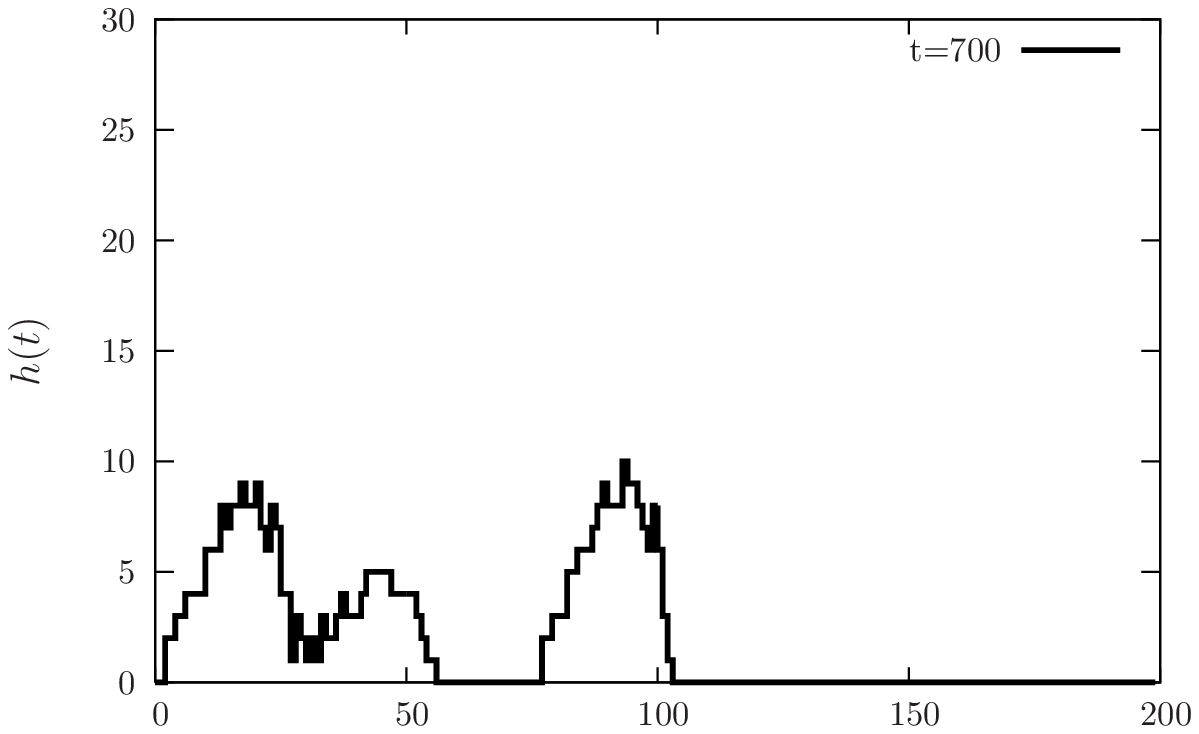}\\
    \includegraphics[width=0.7\columnwidth]{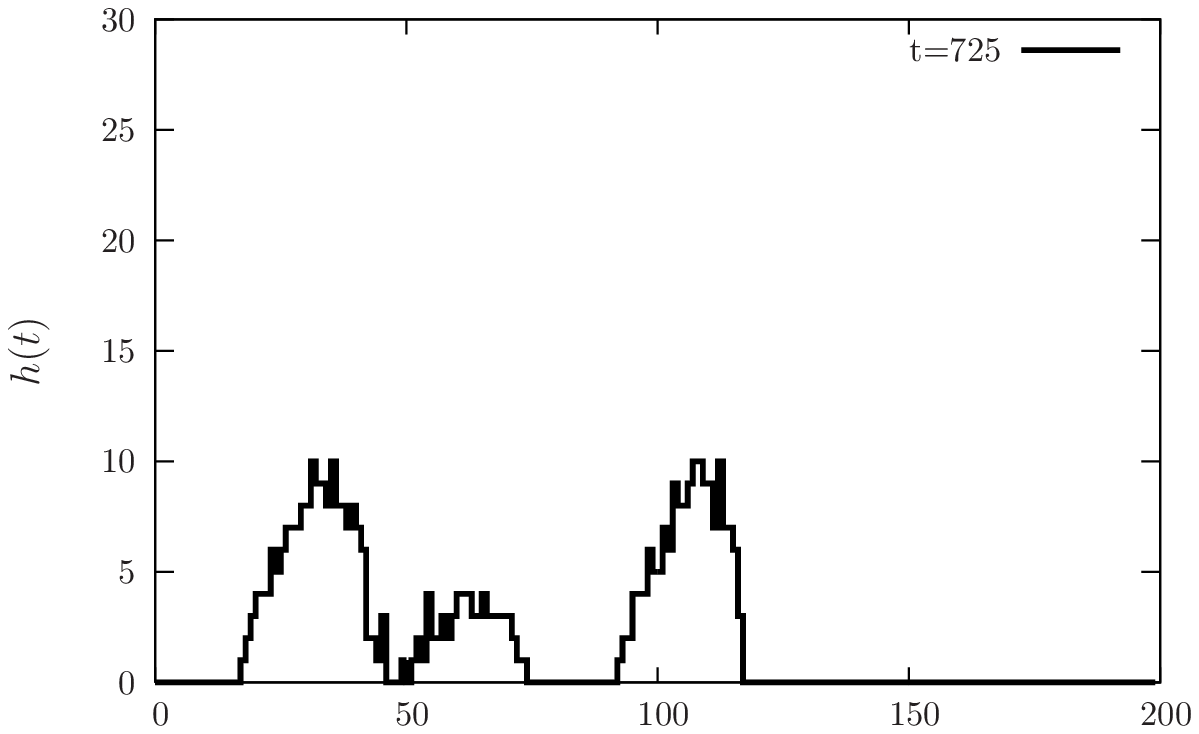}
    \includegraphics[width=0.7\columnwidth]{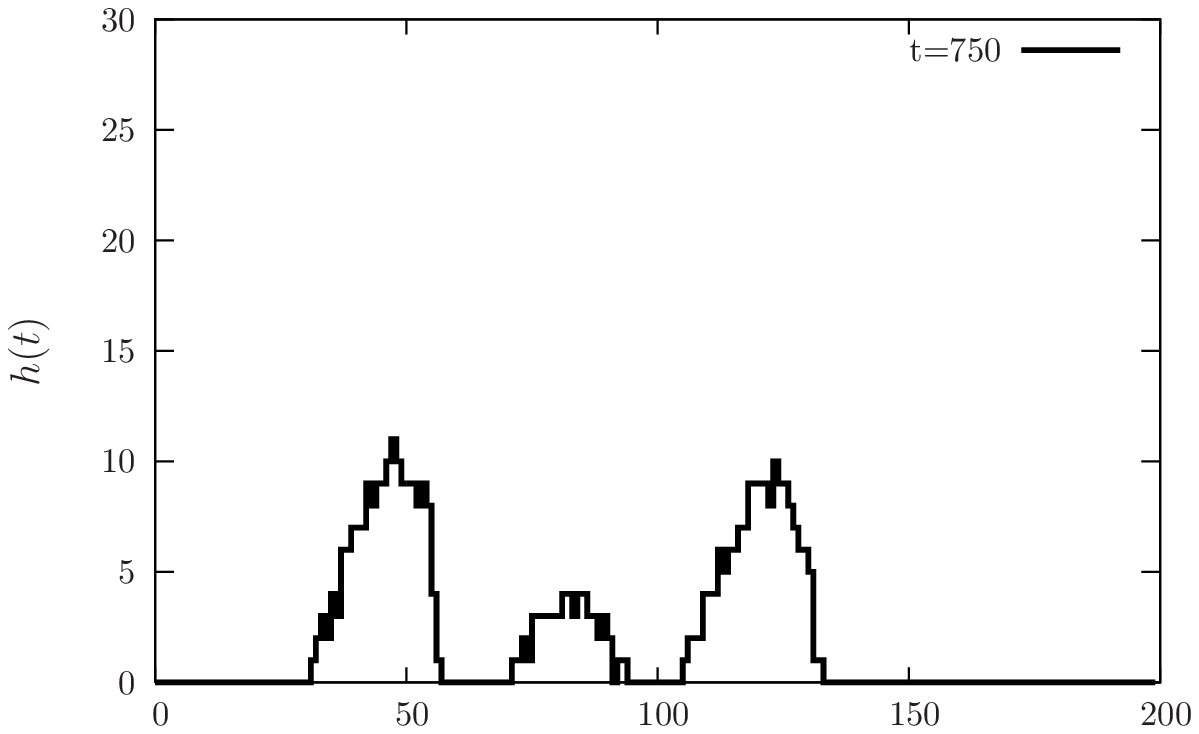}\\
    \caption{When two ripples of similar sizes encounter, an exchange
      in their relative positions occur (take into account periodic
      boundary conditions).\label{fig:exchanging}}
  \end{center}
\end{figure}

\begin{figure}
  \begin{center}
    \includegraphics[width=0.7\columnwidth]{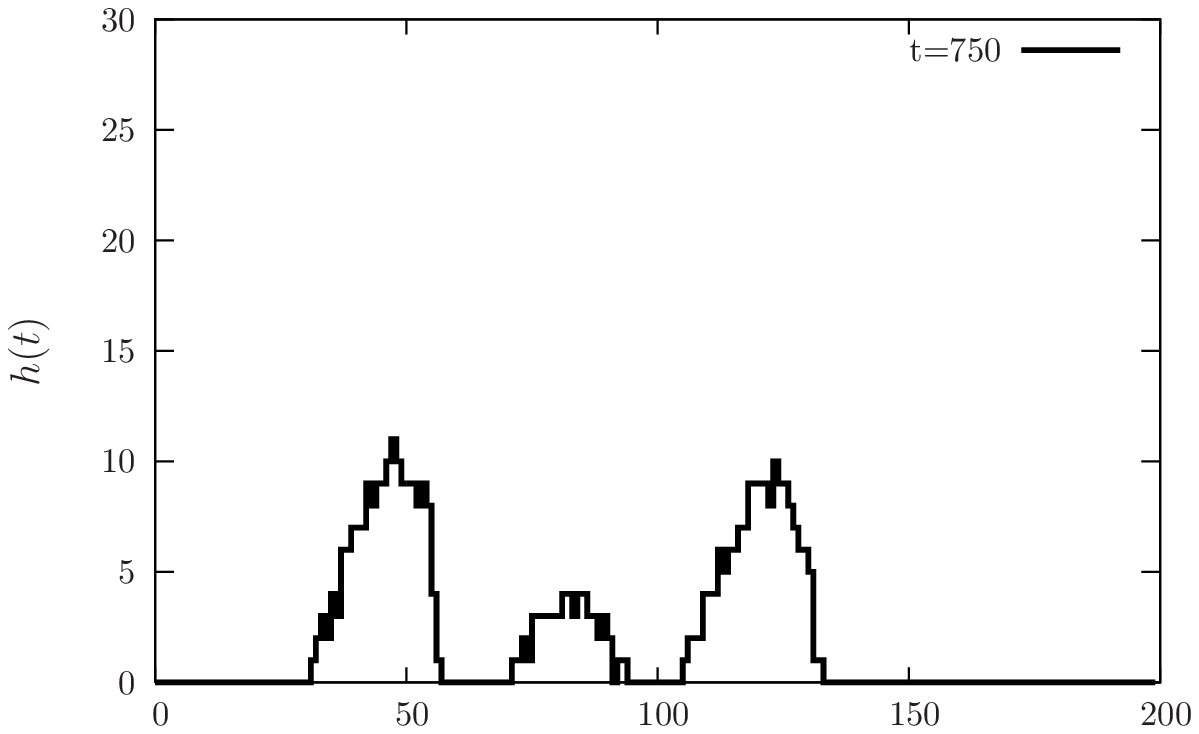}
    \includegraphics[width=0.7\columnwidth]{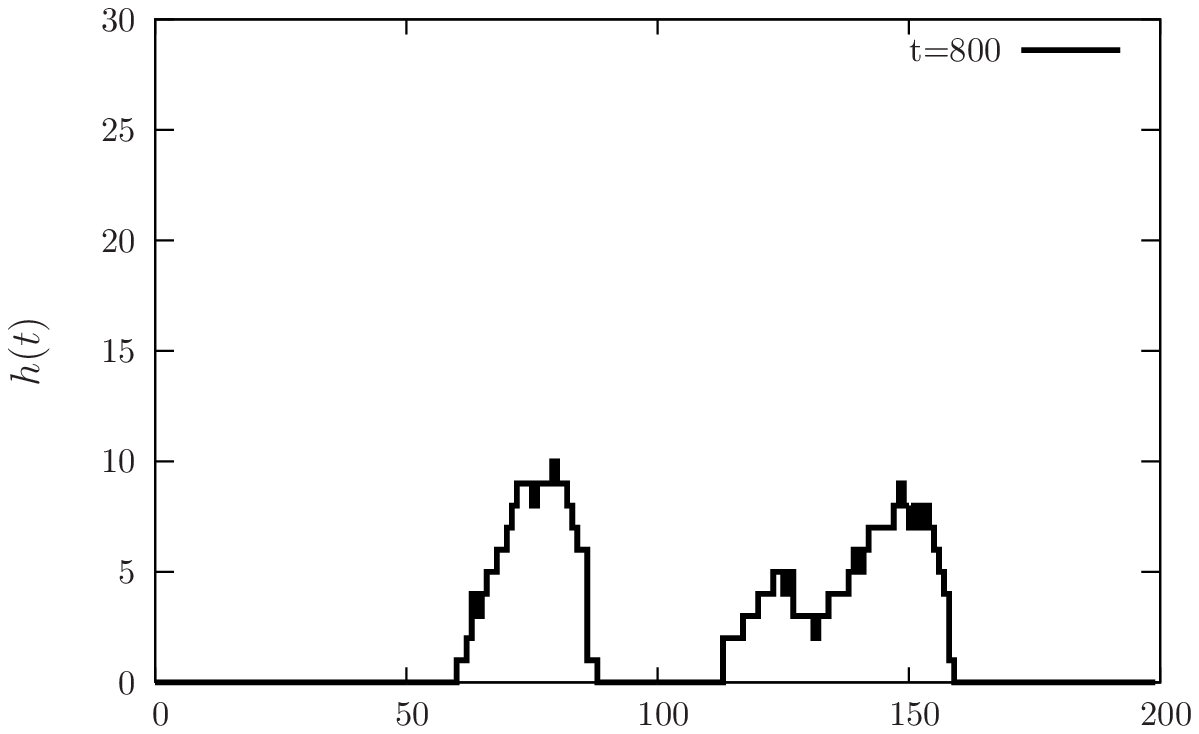}\\
    \includegraphics[width=0.7\columnwidth]{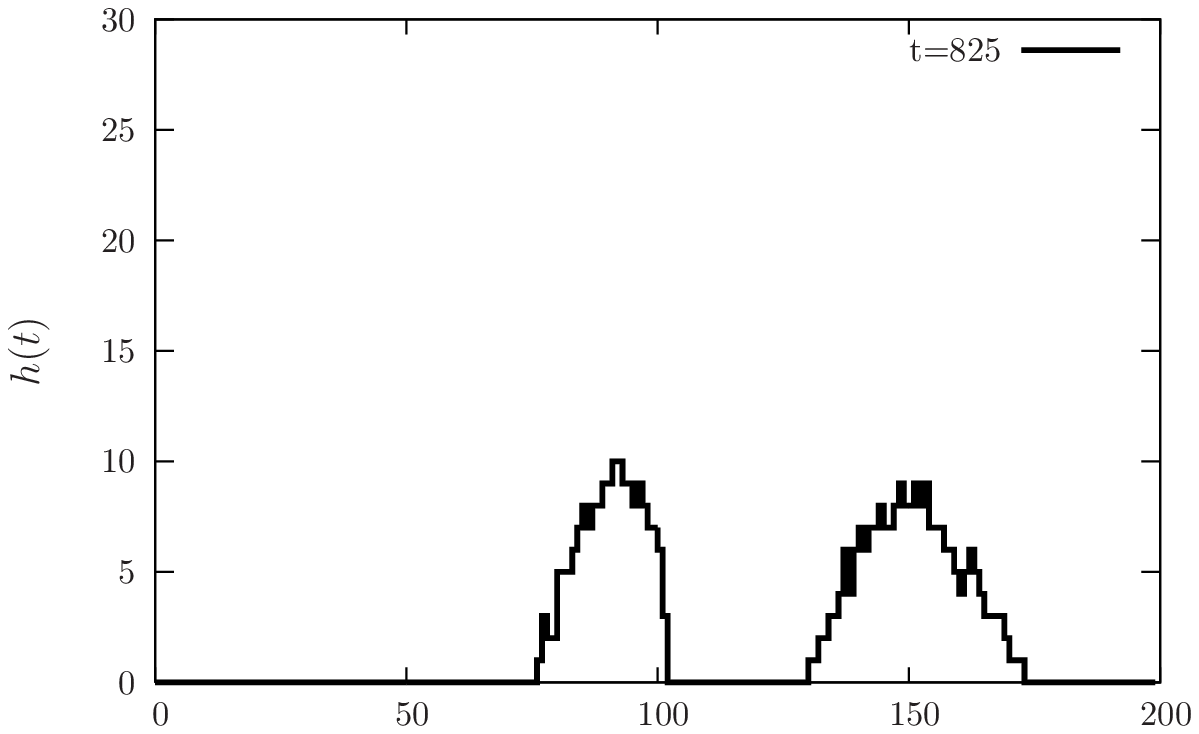}
    \includegraphics[width=0.7\columnwidth]{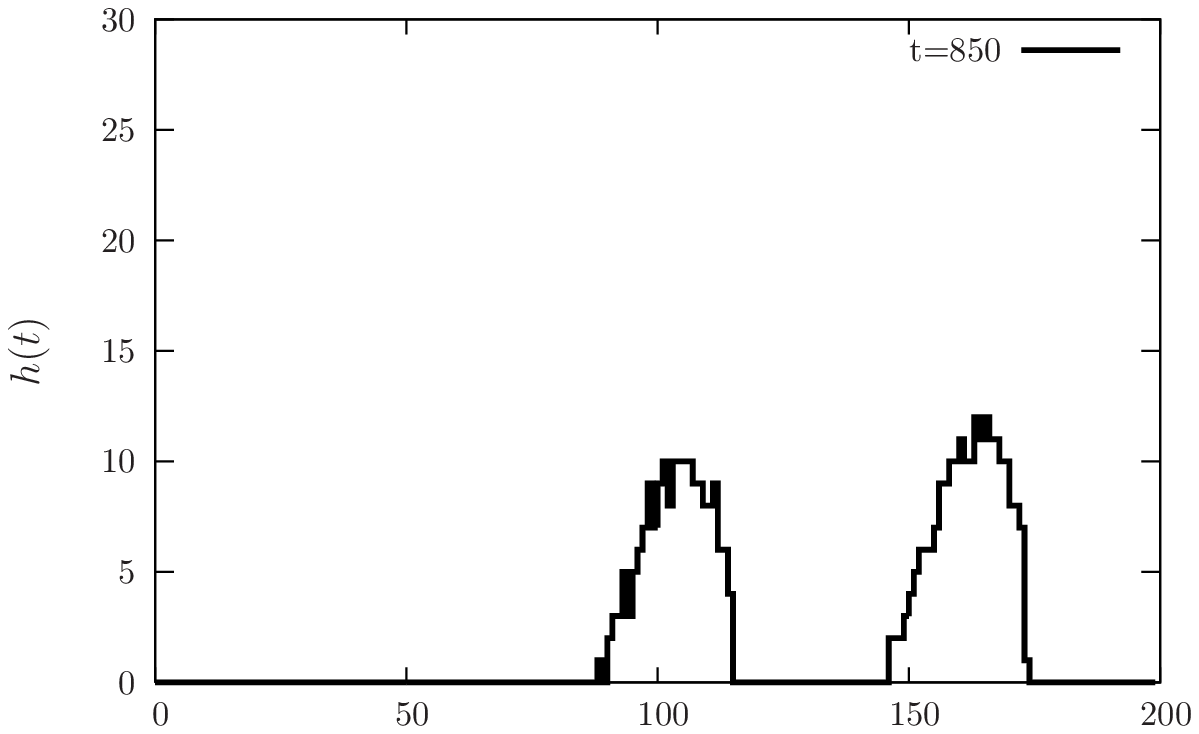}\\
    \caption{An encounter between ripples of different sizes results
      in the merging of the structures.\label{fig:merging}}
  \end{center}
\end{figure}

A sequence of successive snapshots of the dynamics is collected   in
Figure~\ref{fig:evolution} and allows to qualitatively investigate the
process of formation of coherent structures. 

The
displacement of a ripple is a consequence of the combined effects of
erosion and deposition: the grains are eroded in the stoss side and
deposed in the lee one. This is a dynamical mechanism: there is a
continuous exchange between the rest particles and the flowing ones.
The global consequence of this flow is that the lower end of the stoss
side is eroded and this matter deposes in the lee part, leading to the
displacement of the ripple with a velocity that decreases with size. 
  
The interaction mechanism of two ripples is rather complex,
as illustrated by the ``collision'' of two ripples of similar size,
 Fig.~\ref{fig:exchanging}. In order to collide, the front ripple has
to be larger than the rear one, which is consequently faster. 
When the two ripples approach, the lower part
of the stoss side of the front ripple is no more eroded. 
This is due (in the model) to the increasing 
average height, that includes the contribution of 
the approaching rear ripple. 
In the reality, this corresponds to the
reduction in erosion due to the shielding effect of the rear ripple.

The region of the stoss side next to this one becomes the source of eroded
particles, thus forming a local sink. The size of the front ripple is
decreased and its speed increased. If the relative difference of
velocity of the two ripples is low enough (\emph{i.e.} for similar
ripples of similar size) this depression may proceed enough to make the
front ripple detach from the rear one, Fig.~\ref{fig:exchanging}. 

If the rear ripple is small,
the sinking region is continuously moved downwind and
finally the two ripples coalesce. This mechanism is illustrated in
Fig.~\ref{fig:merging}. In the third panel ($t=825$) one can still
recognize a signature of this process: focusing on the right ripple,
originated by the interaction of two ripples of different size (see
first panel), one can still identify the protruding bump on the lee side. This is
the relic of the highest peak displayed by the right ripple  
in the second panel, that experienced a reduction in size and consequently
proceeded faster. However, this bump is eventually screened by the
rear portion of the ripple. As a consequence it stops and is therefore
engulfed in the incoming massive bulk.

These observations are in
agreement with direct measurements and provide a first validation of
our simplified  interpretative framework~\cite{Bagnold}. 

\begin{figure}
  \begin{center}
    \includegraphics[width=1.0\columnwidth]{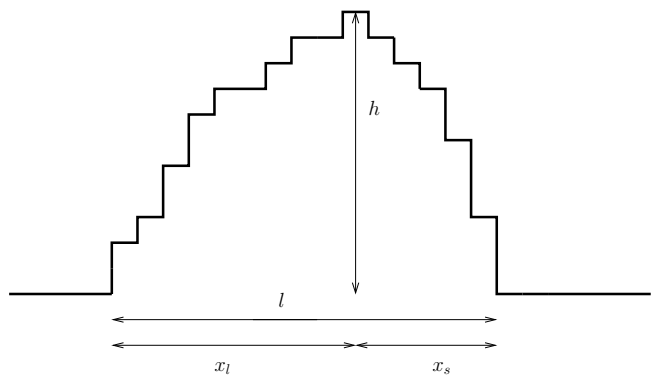} 
    \caption{Ripple schematization, where $x_l$
      ($x_s$) is the projection of the lee (stoss) slope and $h$ is the
      ripple height.\label{fig:ripple-profile}}
  \end{center}
\end{figure}

\begin{figure}
  \begin{center}
    \includegraphics[width=1.0\columnwidth]{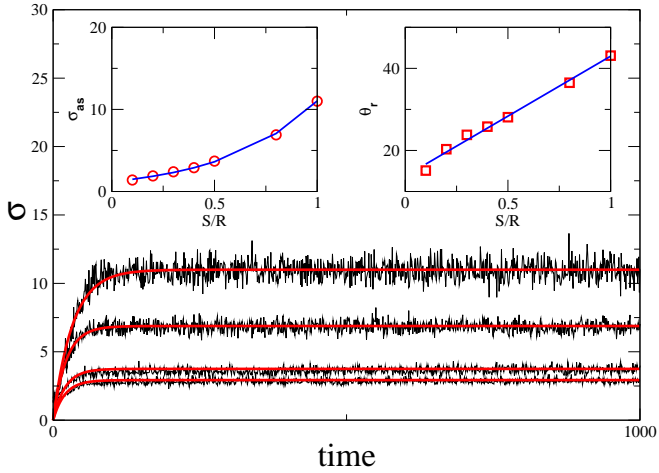} 
    \caption{Time evolution of ripple asymmetry $\sigma$ as a
      function of the ratio $S/R$. Left-inset: asymptotic value of
      ripple asymmetry as a function of $S/R$. Right-inset: plot of the
      dynamic angle of repose $\theta_r$ against the ratio $S/R$
      (circles: numerical values; solid line: numerical
      fit).\label{fig:asim}}
  \end{center}
\end{figure}

\begin{figure}
  \begin{center}
    \vspace{.5cm}
    \includegraphics[width=1.0\columnwidth]{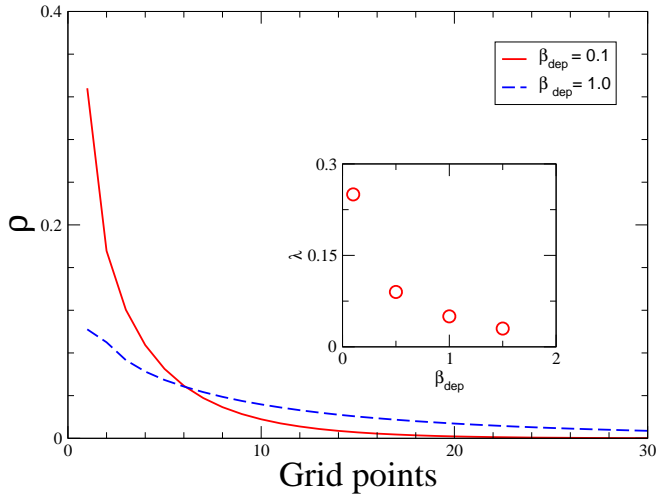}	 
    \caption{Density of particles $\rho$ as a function of the
      deposition distance, for $\beta_{d}=0.1$ and $1.0$. Inset:
      characteristic length scale $\lambda$ of deposition as a function
      of $\beta_{d}$. \label{fig:betadep}}
  \end{center}
\end{figure}

\begin{figure}
  \begin{center}
    \includegraphics[width=1.0\columnwidth]{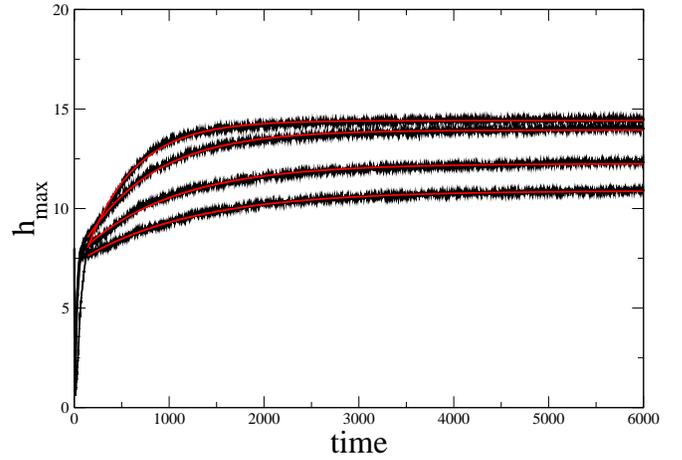} 
    \caption{Exponentially increase of the
      maximum ripple height $h_{max}$ with time for different values of
      the ratio $S/R$. \label{fig:rippleheight}}
  \end{center}
\end{figure}

Label with $l$ the linear size of a typical ripple and assume $h$ to represent its characteristic height. By tuning the parameter  $R$, \emph{i.e.} varying the extension of the segment that defines the interacting region, one modulates the ratio $h/l$ and operates an a priori selection among various types of structures (ripples, megaripples, giantripples) based on their intrinsic geometry. The crucial role of $R$ has been investigated through a dedicated campaign of simulations: by assigning larger values of $R$ corresponds to generate less peaked structure,  which translates into a systematic reduction of the quantitative  indicator $h/l$. Focusing on ripples, we assume $R=32$,  and  speculate on the role of the remaining parameters $S$ and $\beta_{d}$ in  the formation of the ripple~\cite{footnote}: few quantities of paramount importance are monitored and  compared with analogous predictions reported in the classical literature~\cite{NO1993,Caps}. 

As already anticipated, within our simplified scheme, the ripples
present an asymmetric shape which can be measured by introducing the 
\emph{aspect ratio} $\sigma$:

\begin{equation}  
  \sigma = \frac{x_s}{x_l},
\end{equation}  
where $x_s$ (resp. $x_l$) stands for the projection of the stoss
(resp. lee) slope on the horizontal axis, as depicted in
Figure~\ref{fig:ripple-profile}. If $\sigma=1$ the ripples are
symmetric, while $\sigma \ne 0$ implies  an asymmetry. In the main
panel of Figure~\ref{fig:asim}, the evolution of the  ripple aspect
ratio $\sigma$ is plotted as function of time  for different values of
the ratio $S/R$. An initial growth is observed, followed by a
subsequent saturation towards an asymptotic plateau, $\sigma_{as}$.  A
self-consistent selection mechanism is therefore operated by the
system and eventually only one specific class of ripples arises and
occupies the one dimensional lattice. The following ansatz,
\begin{equation}
  \sigma=\sigma_{as} \left( 1- \exp(-\alpha t) \right),
\end{equation}
is numerically fitted to the simulated  profiles of
Figure~\ref{fig:asim}  and shown to interpret well the data. To better
visualize the  tendency of enhancing the degree of asymmetry for
increased values of the local distortion S (working at constant R),
$\sigma_{as}$ is represented versus the ratio $S/R$ in the top-left
inset of Figure~\ref{fig:asim}.  Further, to provide a complete
characterization of the morphology of the ripple, we calculate the
repose angle $\theta_r$, dynamically selected within our proposed
approach, as function of    the control quantity S/R. 

The temporal evolution of $\sigma$ was previously monitored
in~Ref.~\cite{Caps} for both the NO and SCA models. The original NO
formulation predicts almost symmetric profiles and therefore $\sigma
\simeq 1$. Conversely,   for the case of the SCA $\sigma$ grows
linearly in time and then relaxes to a final value. This remarkable
improvement  was achieved by Caps and Vandewalle by postulating the
existence of a repose angle $\theta_r$ and modeling the process of
avalanches, not included in the NO philosophy. It is worth emphasizing
that a somehow similar mechanism (the saturation for $\sigma$ being
here exponentially approached) is  here reproduced without invoking
{\it a priori} the existence of a limiting angle.

In Figure~\ref{fig:betadep} the deposition process is investigated to
shed light into the crucial role of $\beta_{d}$. The  normalized
density of particles $\rho$ as function of the deposition distance is 
plotted for distinct values of $\beta_{d}$.  The measured
distributions decays exponentially and the characteristic length scale
$\lambda$ is represented versus $\beta_{d}$ in the small inset.  As
expected, for larger values of $\beta_{d}$ the particles spend more
time in the surrounding halo and retard the deposition event. We
therefore suggest that  $\beta_{d}$ provides an indirect control of
the characteristic length of the reptation process.

Finally, we studied the dynamical evolution of the maximum ripple height $h_{max}$.  In \cite{Caps} the SCA model was shown to reproduce the non linear evolution of the ripple amplitude $h_{max}$, this success being ascribed to the new ingredients introduced with respect to the NO scenario. Results of our simulations are reported in Figure~\ref{fig:rippleheight}: as for the SCA an exponential growth law is also found in the framework of our probabilistic approach, thus reinforcing its validity as an alternative tool to address the relevant issue of ripple formation.  
 
\section{Conclusion}
\label{sec:conclusion}

The proposed local-equilibrium model for the study of aeolian ripple dynamics has shown to successfully reproduce important observed features,  despite its intrinsic simplicity. In particular, the irreversible and not trivial coarsening dynamics with merging and scattering of structures, the saturating value of the maximum height of the ripples and  the asymmetry of ripple structures have been reproduced. The results are critically compared with the classical literature \cite{NO1993,Caps}, outlining the role of the parameters involved in our formulation and their physical interpretation. 
This simple formulation may be easily extended to include a more detailed description of the fluid flow and the characteristics of the granular phase.


\begin{thebibliography}{00}
\bibitem{footnote} The quantity $\beta_{e}$ refers to the granular material selected and indirectly measure its degree of  packing and internal cohesion. In the present study we assume $\beta_{e}=1$.
\end{thebibliography}
\end{document}